\documentclass[pdflatex,sn-apa]{sn-jnl}

\usepackage{amsmath,amssymb,amsfonts}
\usepackage{bbm}
\usepackage{booktabs}
\usepackage{graphicx}
\usepackage{tabularx}
\usepackage{threeparttable}
\usepackage{makecell}
\usepackage{rotating}
\usepackage{float}
\usepackage{natbib}
\usepackage{xcolor}
\usepackage{hyperref}
\hypersetup{hypertexnames=false}

\begin{document}
	
\title[Socio-temporal effects in microfinance delinquency]{Quantifying socio-temporal effects of loan delinquency drivers in microfinance}

\author*[1]{
	\fnm{Cedric H.A.} \sur{Koffi}\,\href{https://orcid.org/0009-0002-4742-3679}{(ORCID)}}
\email{a.koffi2@liverpool.ac.uk}

\author[2]{
	\fnm{Viani Biatat} \sur{Djeundje}\,	\href{https://orcid.org/0000-0001-9054-2783}	{(ORCID)}}
\email{viani.djeundje@ed.ac.uk}

\author[1]{
	\fnm{Olivier Menoukeu} \sur{Pamen}\,\href{https://orcid.org/0000-0001-9306-3155}{(ORCID)}}
	\email{menoukeu@liverpool.ac.uk}

\affil*[1]{\orgdiv{Institute for Financial and Actuarial Mathematics, Department of Mathematical Sciences},	\orgname{University of Liverpool}, \orgaddress{\city{Liverpool}, \country{UK}}}

\affil[2]{\orgdiv{Credit Research Centre},\orgname{University of Edinburgh Business School}, \orgaddress{\city{Edinburgh}, \country{UK}}}

\abstract{We develop and evaluate a family of discrete-time logit-link (LLink) models, including fixed-effects and frailty extensions, to quantify associations between socio-temporal factors and loan delinquency transitions while accounting for latent borrower heterogeneity. Using monthly records for 1,716 borrowers from a Ghanaian microfinance institution, we model transitions among good, intermediate, and poor repayment states, distinguishing deterioration, persistence, and recovery. The Eid season and long vacation indicators are associated with repayment behaviour, although the magnitude and direction of their estimated effects vary across transitions. Bootstrap tests indicate residual borrower heterogeneity in recovery and persistent poor repayment but provide little support for time-dependent frailty beyond a random intercept. Random Forest and KTBoost perform better for several binary transitions, while the LLink models provide a transparent framework for estimating covariate effects and borrower heterogeneity. For multistate prediction, we propose an Optimised Matthews Correlation Coefficient (OMCC) rule and compare it with the Djeundje and Crook (D\&C) rule. The estimated socio-temporal effects are interpreted as conditional associations rather than causal effects.}

	
\maketitle

\section{Introduction}
The provision of small loans to low-income individuals, commonly referred to as micro-lending, gained prominence following the foundational work of \citet{yunus2007banker}. {By providing credit to people who may have limited access to formal banking, microfinance can support financial inclusion \citep{ledgerwood1999sustainable,armendariz2010economics}}. As noted by \citet{cull2009microfinance} and \citet{d2011women}, microfinance institutions have extended small loans and other financial services to millions of borrowers worldwide, with women forming a large share of their customers.
\medskip

Credit risk remains an important operational concern for these institutions. One difficulty is assessing applicants when formal credit histories or verifiable income are limited \citep{ampountolas2021machine}; another is monitoring repayment after a loan has been approved. A key characteristic of microfinance loans is that repayment is made through small scheduled instalments. During the life of a short loan, a borrower may remain close to schedule, move into partial repayment, become delinquent, or recover before the contract ends. A binary default label does not distinguish these movements.
\medskip

Microfinance credit-scoring studies have used both conventional statistical models and machine learning. \citet{vangool2012credit} examined how logistic credit scoring could support lending decisions, while studies using Peruvian MFI data compared conventional models with neural networks, support vector machines, classification trees, and ensemble methods \citep{blanco2013credit,cubiles2013improving}. 
\medskip

The range of data and methods used in microfinance credit risk studies has also expanded. Recent work covers alternative borrower data, digital credit scores, spatial dependence, and additional classifiers \citep{de2019does, ruiz2019credit, djeundje2021enhancing, ampountolas2021machine, medina2022spatial, dong2023big}. Of particular relevance here, \citet{ampountolas2021machine} included RF among the methods applied to Ghanaian micro-lending data. These studies mainly consider borrower selection or default prediction. Our focus is on monthly movements between repayment states over the life of the loan.
\medskip

Survival models have also been used to study the timing of default and other credit outcomes \citep{banasik1999not,stepanova2002survival,leow2016stability, dirick2022hierarchical}. Multistate models describe movements among credit rating or delinquency states \citep{koopman2008multi,leow2014intensity}. \citet{djeundje2018incorporating} extended this approach by including macroeconomic variables and frailty in a multistate model of credit card delinquency. Discrete-time models are useful for outcomes observed at fixed intervals \citep{singer1993time}. In our data, weekly and fortnightly repayment records are aggregated to monthly intervals. The repayment states constructed at these intervals allow us to distinguish whether a borrower remains in the current state, deteriorates, or recovers.
\medskip

Machine-learning methods, including random forests and boosting, have been widely used for credit scoring and default prediction \citep{brown2012experimental, lessmann2015benchmarking}. We use the LLink models to estimate transition-specific covariate effects and frailty variance components and compare their held-out prediction performance with RF and KTBoost.
\medskip

Repayment may also vary over the year as income, household expenditure, and business conditions change. Studies in Bangladesh have examined seasonal credit use and repayment arrangements designed around uneven cash flows \citep{pitt2002credit, shonchoy2014seasonality}. \citet{weber2017flexible} examined flexible agricultural microfinance loans, while \citet{laureti2017flexible} documented changes in savings and loan repayment around major festive periods.

These studies suggest that repayment behaviour may reflect both seasonal conditions and the social and economic circumstances in which borrowers live and work. Evidence from Ghana shows that social networks can influence how entrepreneurs obtain and use resources \citep{kuada2009gender}, while the seasonal microfinance studies above show that household finances and repayment arrangements may vary over the year. This motivates examining whether local calendar periods are associated with different repayment transitions in the portfolio. In this paper, we focus on the Eid season and long vacation indicators. These periods may coincide with changes in household expenditure, liquidity, or business activity, but the data do not capture those conditions directly; we interpret their estimated effects as conditional associations.
\medskip

Using repayment records from 1,716 borrowers with short-term loans at a microfinance institution in Ghana, we define a three-state repayment process and estimate each modelled movement with a transition-specific discrete-time logit-link (LLink) model. The three states distinguish good, intermediate, and poor repayment, which allows deterioration, persistence, and recovery to be modelled separately. However, borrowers with the same observed characteristics may still differ in their repayment behaviour. Frailty models enable us to assess variation remaining after the observed covariates are included using borrower-level random effects \citep{duchateau2008frailty, hougaard2000frailty, sohn2007random, chamboko2019frailty, dirick2022hierarchical}. In this paper, we estimate fixed-effect LLink models, add random intercepts to assess remaining borrower heterogeneity, and use random linear and random piecewise specifications to examine whether it changes over repayment time.
\medskip

The paper contributes to the analysis of short-term microfinance repayment in three ways. First, the transition-specific LLink models show how the estimated effects of socio-temporal conditions and other observed variables differ across deterioration, persistence, and recovery. Second, the random intercept, random linear, and random piecewise specifications assess where borrower-level heterogeneity remains and whether it changes over repayment time. Third, we propose the Optimised Matthews Correlation Coefficient (OMCC) rule for multistate prediction and compare it with the Djeundje and Crook (D\&C) rule \citep{djeundje2018incorporating}.
\medskip

The paper is structured as follows: Section \ref{section_data_states} presents the data, delinquency states, and transition-types. Section \ref{section_methods} provides the modelling methods. Section \ref{section_results} reports the results, and Section \ref{section_discussion} provides the discussion and conclusion.

\section{Data and Delinquency States}
\label{section_data_states}

\subsection{Portfolio and Covariates}

The empirical analysis uses administrative repayment records from a microfinance institution in Ghana. After removing incomplete records and records for which the repayment state could not be constructed reliably, the working portfolio contains 1,716 borrowers with complete transaction histories, where each borrower has one loan in the study period. The loans were observed between April 2018 and November 2018, and all had repayment periods shorter than eight months. Repayment was scheduled monthly, weekly, or fortnightly, with monthly instalments being the most common schedule. Weekly and fortnightly repayment records are aggregated to the monthly level before the borrower-period transitions are constructed. We focus on borrower-period transitions because the data provide repayment information at monthly intervals. A one-period transition compares a borrower’s repayment state in one month with the state in the following month.
\medskip

The socio-temporal variables were constructed from calendar periods in the observation window. The lagged macroeconomic variables were obtained from Ghana Statistical Service StatsBank, the official data dissemination platform \citep{ghanaStatsBank}.

 Because the data cover only April to November 2018, the calendar indicators, lagged macroeconomic variables, and loan duration may capture overlapping time patterns. We therefore interpret their estimated effects as conditional associations rather than causal effects.
\medskip

The data sharing agreement does not allow us to report some portfolio summaries, including the mean loan size and the distribution of interest rates. Table \ref{table_compact_variable_description} therefore describes the variables used in the models without reporting full descriptive statistics. Eid season and long vacation are calendar indicators; the data do not capture the household or business conditions that may explain their estimated effects.

\begin{table}[!h]
	\centering
	\caption{Variables used in the transition models}
	\label{table_compact_variable_description}
	\begin{tabularx}{\textwidth}{p{0.24\textwidth}X}
		\toprule
		Variable group & Definition \\
		\midrule
		Repayment history &
		{Two month lagged indicator that the borrower had accumulated at least two previous partial repayments.} \\
		
		Socio-temporal indicators &
		{Eid season is coded 1 in June and August 2018, and long vacation is coded 1 in July and August 2018; both are coded 0 otherwise.} \\
		
		Loan characteristics &
		{Principal, interest rate, repayment frequency, group loan indicator, and branch indicator. Monthly repayment is coded 1 for a monthly schedule and 0 otherwise; group loan is coded 1 for group lending and 0 otherwise; main branch is coded 1 and other company branches are coded 0.} \\
		
		Borrower characteristics &
		{Age group indicators, gender, and marital status. The age groups, adapted from \citet{silinskas2021financial}, are 26--35, 36--45, 46--55, and 56 or older, with borrowers aged 25 or younger as the reference group. Female is coded 1, and married is coded 1.} \\
		
		Macroeconomic variables &
		Third monthly lags of CPI and the USD/GHS exchange rate. \\
		\bottomrule
	\end{tabularx}
	
	\begin{tablenotes}
		\footnotesize
		\item {Principal, CPI, and the USD/GHS exchange rate are min--max scaled in the modelling data.}
	\end{tablenotes}
\end{table}

\subsection{Repayment States}
The state definition is built from the fact that partial repayment is common in the transaction records of the company. Borrowers often pay part of the scheduled amount across multiple periods, and the institution usually {continues} to manage the loan rather than assigning it immediately to an absorbing or default state. We define repayment states from the ratio of the amount paid in the current period to the amount scheduled for that period. Let $k_i(t)$ denote the amount repaid by borrower $i$ at time $t$, and let $A_i(t)>0$ denote the scheduled repayment amount. The current-period repayment ratio is
$$
  R_i(t)=\frac{k_i(t)}{A_i(t)}.
$$
The three repayment states are defined as
$$
S_i(t)=
\begin{cases}
1, & R_i(t)\geq 0.82 \quad \text{good repayment},\\ 2, & 0.60\leq R_i(t)<0.82 \quad \text{intermediate repayment},\\ 3, & 0\leq R_i(t)<0.60 \quad \text{poor repayment}.
\end{cases}
$$
A borrower is treated as delinquent at time $t$ when the loan is in state 2 or state 3. These states classify the repayment made against the scheduled amount for the current period; repayment history is included through a two month lagged partial repayment indicator.
\medskip

The thresholds were chosen in consultation with the partner institution and assessed through the threshold-sensitivity analysis reported in Appendix~\ref{section_diagnostics}. Figure~\ref{figure_compact_mean_trajectories} shows the average repayment ratio at each repayment period, conditional on the borrower’s origin state in the preceding period. Since all borrowers are in state 1 before their first repayment, transitions from states 2 and 3 are observed only from the second repayment period. The three-state construction reflects the repayment patterns observed in this portfolio and distinguishes different degrees of incomplete repayment. Partial repayment has also been documented in another microfinance setting \citep{mirpourian2016determinants}.

\begin{figure}[H]
	\centering
	\includegraphics[width=0.86\textwidth]{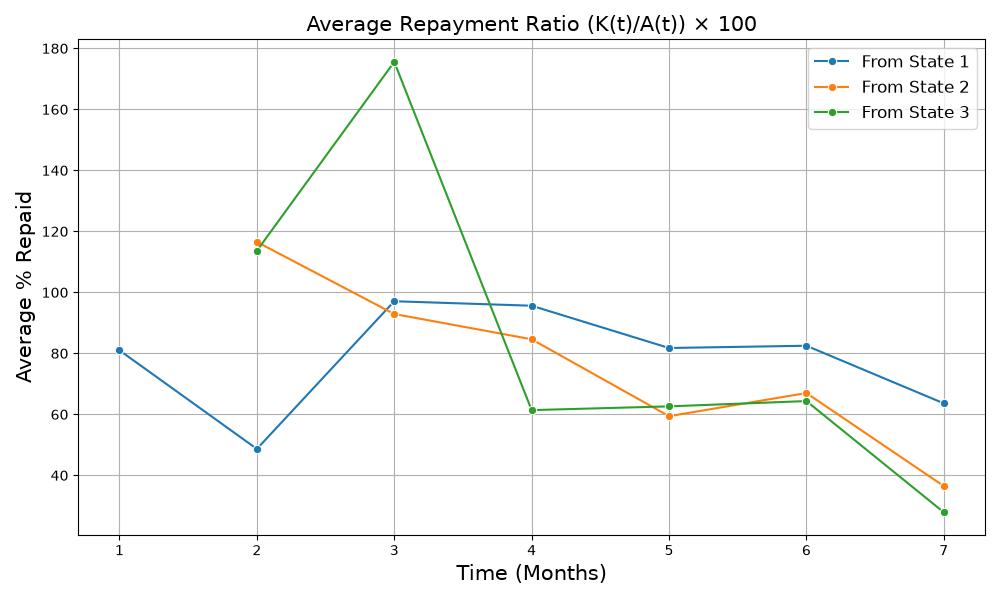}
	\caption{Average repayment ratio by origin state and repayment period.}
	\label{figure_compact_mean_trajectories}
\end{figure}
{We do not impose an absorbing default state because borrowers in poor repayment can later recover in the observed data. The loans are short and partial repayments are frequent, so an absorbing default label would hide some of the repayment movement that we study.}

	\subsection{Transition-types}

Time is measured as duration (in months) since loan disbursement rather than calendar time. This puts borrowers on a common repayment clock, while still allowing calendar-based covariates, such as Eid season and long vacation, to vary across repayment periods. Loans are scheduled at different repayment frequencies, with monthly instalments being the most common. We account for these schedule differences using a monthly-repayment indicator, coded 1 for monthly schedules and 0 for non-monthly schedules, rather than estimating separate transition models for each repayment schedule. This also avoids splitting the data into smaller schedule-specific samples and helps stabilise the coefficient estimates of the modelled transition probabilities.
\medskip

With three repayment states, nine one-period transitions are possible, but in the main empirical models, we focus on the following six sub-models:
$$
\mathcal T=\{(1,1),(1,3),(2,1),(2,3),(3,1),(3,3)\}.
$$
These transitions cover persistence in good repayment, deterioration into poor repayment, recovery from intermediate or poor repayment, deterioration from intermediate repayment, and persistence in poor repayment. We model the two destinations, states 1 and 3, directly. The intermediate-state probability is obtained as the residual for multistate prediction; state 2 is also the least common destination overall. Table~\ref{table_compact_transition_support} reports the observed transition counts by origin and destination state.

\begin{table}[!h]
\centering
\caption{Count of observed transitions from origin to destination state}
\label{table_compact_transition_support}
\footnotesize
\begin{tabular}{lrrrr}
\toprule
Origin state & Destination 1 & Destination 2 & Destination 3 & Origin rows \\
\midrule
State 1 & 3,454 & 789 & 1,820 & 6,063 \\
State 2 & 369 & 100 & 260 & 729 \\
State 3 & 1,296 & 126 & 714 & 2,136 \\
\midrule
Total & 5,119 & 1,015 & 2,794 & 8,928 \\
\bottomrule
\end{tabular}
\end{table}

For each modelled transition $(h,j)$, the response is defined on borrower-period rows that start in origin state $h$. It is equal to one when the borrower moves to destination state $j$, and zero when the borrower moves from state $h$ to any other observed destination state. {The sub-models are modelled separately, so each fitted equation gives transition-specific parameter estimates and event probabilities for the selected destination.} When a full three-state probability vector is needed, the destination-2 probability is computed as the residual
$$
\hat q_{i,h2}(t)=1-\hat q_{i,h1}(t)-\hat q_{i,h3}(t),
$$
where $\hat q_{i,hj}(t)$ is the estimated transition probability from state $h$ to state $j$ at time $t$\footnote{Because $\hat q_{i,h1}(t)$ and $\hat q_{i,h3}(t)$ are estimated from separate binary transition models, their sum may occasionally exceed one and produce a negative residual. In those cases, the vector is adjusted and renormalised so that all three entries are non-negative and sum to one.}.

The three-state probability vectors $(\hat q_{i,h1}(t),\hat q_{i,h2}(t),\hat q_{i,h3}(t))$ are used for multistate next-state prediction and threshold-classification summaries. The corresponding threshold rule and results are reported in Section~\ref{subsec_results_prediction_checks}.

\section{Modelling Approach}
\label{section_methods}

{The models use the transition structure defined in the previous section. For borrower $i$, let $S_i(t)\in\{1,2,3\}$ denote the repayment state at monthly repayment period $t$, and let $T_i$ be the borrower's last observed period. {Let $t_0=0,t_1=1,\ldots,t_M=T$ denote the discrete monthly repayment-duration grid, where $t_m=m$ and $T=\max_i \{T_i\}$.} {At $t_0$, no repayment has occurred, so $S_i(t_0)=1$ for every borrower. The first repayment state is observed at $t_1$, and transitions from origin states 2 and 3 can therefore occur only from the second repayment month onwards.} Let $N$ be the number of borrowers. Borrower $i$ is at risk of transitioning from origin state $h$ at period $t-1$ when $S_i(t-1)=h$. The corresponding risk set is}
{
$$
\mathcal R_h(t-1)=\{i\in\{1,\ldots,N\}:S_i(t-1)=h\}.
$$
}
{The likelihood function is built from borrower-period transitions for which both the origin state $S_i(t-1)$ and destination state $S_i(t)$ are available.} For a modelled transition $(h,j)\in\mathcal T$, define
$$
Y_{i,hj}(t)=\mathbbm{1}_{\{S_i(t-1)=h,\; S_i(t)=j\}}.
$$
The corresponding conditional transition probability is
$$
q_{i,hj}(t)=
\mathbb{P}\big(S_i(t)=j\mid S_i(t-1)=h,\boldsymbol X_{i,t}\big),
$$
where $\boldsymbol X_{i,t}$ contains the borrower, loan, macroeconomic, repayment-history, and socio-temporal covariates observed before the transition.
\medskip

The fixed-effect logit-link (LLink) model relates the transition probability to a duration effect and the observed covariates:
\begin{align}	\label{eq_compact_fixed_llink}
	\text{logit}\big(q_{i,hj}(t)\big)	= 	\alpha_{hj,t}+\boldsymbol X_{i,t}^{\top}\boldsymbol\beta_{hj},
\end{align}
where $\alpha_{hj,t}$ is a transition-specific duration effect and $\boldsymbol\beta_{hj}$ is the vector of transition-specific covariate coefficients.

Other formulations are possible, including multinomial transition models \citep{agresti2013categorical}. We use a destination-versus-rest formulation because the analysis focuses on transition-specific covariate signs, frailty variance components, and binary likelihood-ratio tests for each modelled {transition-type}.

\medskip

We add a transition-specific random intercept to capture borrower-level variation that remains after accounting for the observed covariates and duration effects:
\begin{align}\label{eq_compact_random_intercept}
	\text{logit}\big(q_{i,hj}(t\mid u_{i,hj})\big) = \alpha_{hj,t}+\boldsymbol X_{i,t}^{\top}\boldsymbol\beta_{hj}+u_{i,hj},	\qquad
	u_{i,hj}\sim N(0,\sigma^2_{u,hj}).
\end{align}

As specification checks, we also estimate two time-dependent frailty extensions that allow the unobserved borrower effect to vary over the repayment period. The time-dependent frailty terms provide the repayment duration structure in these extensions, so a separate duration baseline is not included. The first extension represents the borrower-specific unobserved effect as a linear function of repayment duration:
\begin{align}\label{eq_compact_random_linear}
	\text{logit}\big(q_{i,hj}(t\mid a_{i,hj},b_{i,hj})\big)&= \boldsymbol X_{i,t}^{\top}\boldsymbol\beta_{hj}	+a_{i,hj}t+b_{i,hj}, \\	(a_{i,hj},b_{i,hj})^{\top}
	&\sim	N\left[ \boldsymbol 0, \begin{pmatrix} \sigma^2_{a,hj} & 0 \\ 0 & \sigma^2_{b,hj}
	\end{pmatrix} \right].
	\notag
\end{align}
Within each transition-specific model, the random slope $a_{i,hj}$ and random intercept $b_{i,hj}$ are assumed to have mean zero and to be independent. %Their respective variances are $\sigma^2_{a,hj}$ and $\sigma^2_{b,hj}$.
\medskip

The second LLink extension divides the seven modelled repayment period grid points into three sets 
\begin{equation}
	\mathfrak T_1=\{1,2,3\}, \quad \mathfrak T_2=\{4,5\}, \quad \text{and } \quad \mathfrak T_3=\{6,7\}.\label{defRP1}
	\end{equation} These sets represent the early, middle, and later repayment periods, respectively. For $t\in\mathfrak T_k$, a separate borrower-specific random effect is used and remains constant within repayment period group $k$. In this specification, the model is given as
\begin{align}\label{eq:compact-random-piecewise}
	\text{logit}\big(q_{i,hj}(t\mid u_{i,hj,k})\big) &=	\boldsymbol X_{i,t}^{\top}\boldsymbol\beta_{hj}	+u_{i,hj,k}, \\	u_{i,hj,k}
	&\sim N(0,\sigma^2_{hj,k}),\qquad k=1,2,3.	\notag
\end{align}
where $u_{i,hj,k}$ is the borrower-specific random effect for transition $(h,j)$ during repayment period group $k$, and $\sigma^2_{hj,k}$ is its variance. Within each transition-specific model, the three repayment period-specific random effects are also assumed to be independent. Across all frailty models, the random effects are assumed to be independent across borrowers and independent of the observed covariates.

The two extensions examine whether unobserved borrower heterogeneity changes over the repayment period. Their variance component results and coefficient estimates are reported in Section \ref{subsec_results_latent_heterogeneity}, and their prediction performance is reported in Section \ref{subsec_results_prediction_checks}.
\medskip

Conditional on the origin state, repayment history, observed covariates, duration effects, and any transition-specific frailty term, the borrower-period likelihood contributions are assumed to be independent. The coefficient and frailty results in Section \ref{section_results} are based on these transition-specific binary likelihoods.
\medskip

{For each borrower $i$ and origin state $h$, define the exposure times as}
{
$$
\begin{aligned}
I_{i,h} &=\{t\in\{t_1,\ldots,T_i\}:i\in\mathcal R_h(t-1)\}\\ &=\{t\in\{t_1,\ldots,T_i\}:S_i(t-1)=h\},
\end{aligned}
$$
}
{where $I_{i,h}$ indexes the monthly periods in which borrower $i$ is at risk of transitioning from origin state $h$.}

For the fixed-effect (FE) model, the likelihood for transition $(h,j)$ is the product of Bernoulli terms
\begin{align*}%	\label{eq:compact-fixed-likelihood}
	L^{FE}_{hj}({\boldsymbol\theta^{FE}_{hj}}) = \prod_i\prod_{t\in I_{i,h}} q_{i,hj}(t)^{Y_{i,hj}(t)}
	\big(1-q_{i,hj}(t)\big)^{1-Y_{i,hj}(t)} ,
\end{align*}
where ${\boldsymbol\theta^{FE}_{hj}}$ is the collection of duration and covariate parameters for transition $(h,j)$.
\medskip

For the random intercept (RI) LLink model \eqref{eq_compact_random_intercept}, define the borrower-specific conditional likelihood contribution
$$
f^{RI}_{i,hj}(u):= \prod_{t\in I_{i,h}}q_{i,hj}(t\mid u)^{Y_{i,hj}(t)} \big(1-q_{i,hj}(t\mid u)\big)^{1-Y_{i,hj}(t)},
$$
where $f^{RI}_{i,hj}(u)$ is the product of the Bernoulli likelihood contributions for borrower $i$, conditional on the random intercept $u$. The marginal likelihood is

{\small
	\begin{align}  \label{eq_compact_ri_likelihood}
  L^{RI}_{hj}({\boldsymbol\theta^{RI}_{hj}}) =  \prod_i \int_{\mathbb{R}} f^{RI}_{i,hj}(u)  g(u;0,\sigma^2_{u,hj})\,\mathrm{d}u,
\end{align}}
where $g(u;0,\sigma^2_{u,hj})$ is the Gaussian density with mean $0$ and variance $\sigma^2_{u,hj}$, and {$\boldsymbol\theta^{RI}_{hj}$ contains the duration, covariate, and random intercept variance parameters}. Integrating over the random effect gives a marginal likelihood for {transition-type} $(h,j)$ and allows us to estimate the variance component $\sigma^2_{u,hj}$. Testing whether this variance component is different from zero then checks whether there is statistically significant borrower-level variation remaining after accounting for the observed covariates and duration effects. 
\medskip

The integral in Equation \eqref{eq_compact_ri_likelihood} is evaluated by Gauss-Hermite quadrature (GHQ) \citep{liu1994note}. If $z_r$ and $w_r$ are the Hermite nodes and weights and $Q$ is the number of quadrature points, the {random intercept} marginal likelihood is approximated by
\begin{align*}%\label{eq:compact-ri-ghq}
	L^{RI}_{hj}({\boldsymbol\theta^{RI}_{hj}}) \approx \prod_i\sum_{r=1}^{Q}\frac{w_r}{\sqrt{\pi}}	f^{RI}_{i,hj}(\sqrt{2}\sigma_{u,hj}z_r).
\end{align*}
We use $Q=20$ quadrature points in all three GHQ approximations. For the random linear (RL) model, let
$$
f^{RL}_{i,hj}(a,b)= \prod_{t\in I_{i,h}}q_{i,hj}(t\mid a,b)^{Y_{i,hj}(t)} \big(1-q_{i,hj}(t\mid a,b)\big)^{1-Y_{i,hj}(t)}.
$$
The function $f^{RL}_{i,hj}(a,b)$ represents the borrower-specific conditional likelihood contribution given random slope $a$ and random intercept $b$. The marginal likelihood and its two-dimensional product-rule GHQ approximation are
\begin{align*}
L^{RL}_{hj}({\boldsymbol\theta^{RL}_{hj}}) &=\prod_i\int_{\mathbb R}\int_{\mathbb R} f^{RL}_{i,hj}(a,b)g(a;0,\sigma^2_{a,hj})g(b;0,\sigma^2_{b,hj})\,\mathrm{d}a\,\mathrm{d}b\\
&\approx\prod_i\sum_{r=1}^{Q}\sum_{s=1}^{Q} \frac{w_rw_s}{\pi} f^{RL}_{i,hj}(\sqrt{2}\sigma_{a,hj}z_r, \sqrt{2}\sigma_{b,hj}z_s).
%\label{eq:compact-rl-ghq}
\end{align*}
Here, $\boldsymbol\theta^{RL}_{hj}$ contains the covariate coefficients and the random slope and random intercept variance parameters.
\medskip

For the random piecewise (RP) model, define the repayment period-specific exposure set
$$
I_{i,h,k}:=I_{i,h}\cap\mathfrak T_k,
$$
where  $\mathfrak T_k$ is given in \eqref{defRP1} and $I_{i,h,k}$ contains the exposure periods for borrower $i$ in origin state $h$ that belong to repayment period group $k$. Define
$$
f^{RP}_{i,hj,k}(u_k):= \prod_{t\in I_{i,h,k}}q_{i,hj}(t\mid u_k)^{Y_{i,hj}(t)} \{1-q_{i,hj}(t\mid u_k)\}^{1-Y_{i,hj}(t)},
$$
where $f^{RP}_{i,hj,k}(u_k)$ is the borrower-specific conditional likelihood contribution for repayment-period group $k$, given its group-specific random effect $u_k$. Assuming that the three group-specific random effects are independent, the random piecewise marginal likelihood is
\begin{align*}
L^{RP}_{hj}({\boldsymbol\theta^{RP}_{hj}}) &=\prod_i\prod_{k=1}^{3}\int_{\mathbb R} f^{RP}_{i,hj,k}(u_k)g(u_k;0,\sigma^2_{hj,k})\,\mathrm{d}u_k \notag\\
&\approx\prod_i\prod_{k=1}^{3}\sum_{r=1}^{Q} \frac{w_r}{\sqrt{\pi}} f^{RP}_{i,hj,k}(\sqrt{2}\sigma_{hj,k}z_r). %\label{eq:compact-rp-ghq}
\end{align*}
where $\boldsymbol\theta^{RP}_{hj}$ contains the covariate coefficients and the three repayment period-specific variance parameters. By convention, the product equals one when $I_{i,h,k}$ is empty. An alternative estimation framework for the model parameters using the expectation-maximisation \citep{mclachlan2008algorithm} as well as further details of these frailty formulations are provided in \citet{koffi2026dependence}.
\medskip

Under the null hypothesis $H_0:\sigma^2_{u,hj}=0$, the random intercept variance lies on the boundary of the parameter space. This makes the usual chi-square reference distribution for the likelihood ratio statistic {inappropriate} in this setting \citep{self1987asymptotic,stram1994variance,crainiceanu2004likelihood}. We therefore assess evidence for a nonzero random intercept variance using a parametric bootstrap likelihood-ratio test (LRT) \citep{davison1997bootstrap}. For {transition-type} $(h,j)$, the observed statistic is
\begin{align}
	\label{eq:compact-frailty-lrt}
	\Lambda_{hj} = 2\big(	\ell^{RI}_{hj}(\widehat{\boldsymbol\theta}^{RI}_{hj})	-	\ell^{FE}_{hj}(\widehat{\boldsymbol\theta}^{FE}_{hj})	\big),
\end{align}
where $\ell^{RI}_{hj}$ and $\ell^{FE}_{hj}$ represent the maximised log-likelihoods for {transition-type} $(h,j)$ under the random intercept and fixed-effect models, respectively.

\medskip

The full frailty model estimates are reported in Section \ref{subsec_results_latent_heterogeneity}, while the parametric bootstrap procedure and the remaining diagnostic and sensitivity checks are reported in Appendix~\ref{app_variance_tests}.

Covariate $p$-values in all four estimate tables are Wald-type $p$-values based on the estimated Hessian \citep{agresti2013categorical}, while the variance component $p$-values are from the parametric bootstrap LRTs.

\section{Results}
\label{section_results}

Figure~\ref{figure_compact_mean_trajectories} and the observed transition counts show that repayment patterns differ by origin state. We therefore estimate separate models for persistence in good repayment $(1,1)$, deterioration from good to poor repayment $(1,3)$, recovery from intermediate or poor repayment $(2,1)$ and $(3,1)$, deterioration from intermediate repayment $(2,3)$, and persistence in poor repayment $(3,3)$.

\subsection{{Estimated effects of socio-temporal and other covariates}}

Table~\ref{tab:compact-selected-effects} reports log-odds coefficients from {the separate fixed-effect LLink sub-model for each {transition-type} $(h,j)$}. A positive coefficient increases the conditional log-odds of that transition, while a negative coefficient reduces it, holding the other variables in the same model fixed. The table provides selected fixed-effect estimates for the covariates discussed in this section. {Tables~\ref{table_full_fe_estimates} to \ref{table_rp_estimates}, in Section \ref{subsec_results_latent_heterogeneity}, report the full covariate estimates and frailty standard deviations.}

\begin{table}[!h]
\centering
\caption{Selected fixed-effect LLink estimates by transition-type}
\label{tab:compact-selected-effects}
\scriptsize
\renewcommand{\arraystretch}{0.80}
\begin{tabularx}{\textwidth}{p{0.19\textwidth} *{6}{>{\centering\arraybackslash}X}}
\toprule
Covariate & (1,1) & (1,3) & (2,1) & (2,3) & (3,1) & (3,3) \\
\midrule
Principal & $-2.19^{**}$ & $2.60^{***}$ & $-1.26$ & $-0.52$ & $-2.73^{*}$ & $1.57$ \\
CPI (lag) & $1.79^{***}$ & $-3.16^{***}$ & $-2.03^{*}$ & $2.00^{*}$ & $-4.54^{***}$ & $4.71^{***}$ \\
FX (lag) & $0.34$ & $-0.99^{**}$ & $0.49$ & $0.03$ & $0.92^{*}$ & $-1.44^{***}$ \\
Partial repay. (lag) & $-2.52^{*}$ & $2.40^{***}$ & $-0.31$ & $1.16^{*}$ & $-1.17^{*}$ & $0.33$ \\
Long vacation & $1.07^{***}$ & $-1.37^{***}$ & $-0.03$ & $0.59^{\dagger}$ & $-1.24^{***}$ & $1.04^{***}$ \\
Eid season & $1.07^{***}$ & $-1.38^{***}$ & $-0.44$ & $0.15$ & $0.24$ & $-0.25$ \\
Group loan & $-0.27^{**}$ & $0.19^{\dagger}$ & $0.06$ & $-0.31$ & $0.19$ & $-0.23$ \\
Monthly & $0.54^{***}$ & $0.15$ & $-0.37$ & $0.27$ & $0.93^{***}$ & $-0.77^{***}$ \\
\bottomrule
\end{tabularx}

\begin{tablenotes}
\footnotesize
\item $^{\dagger}p<0.10$, $^{*}p<0.05$, $^{**}p<0.01$, $^{***}p<0.001$. Coefficients are from the fixed-effect model in Equation~\eqref{eq_compact_fixed_llink}.
\end{tablenotes}
\end{table}

{The socio-temporal associations are clearest for borrowers who start the period in good repayment. Eid season and long vacation have the same sign pattern for the two transition-types with origin state 1. Both are positively associated with remaining in good repayment and negatively associated with moving directly from good to poor repayment. These calendar indicators do not capture the household or business conditions that may explain their estimated effects.}
\medskip

For borrowers already in poor repayment, however, long vacation is negatively associated with recovery to good repayment and positively associated with remaining in poor repayment. This contrasts with its associations for borrowers starting in state 1.

\medskip

For the macroeconomic variables, lagged CPI is statistically significant in all six fixed-effect transition models. Higher lagged CPI is associated with higher odds of remaining in good repayment and lower odds of moving directly from good to poor repayment. For transitions starting from states 2 and 3, the pattern is different. Higher lagged CPI is associated with lower recovery to good repayment and higher persistence in poor repayment. The FX lag is statistically significant in fewer transitions, mainly for deterioration from good repayment, recovery from poor repayment, and persistence in poor repayment.
\medskip

Repayment-history and loan-term variables also show clear transition-specific patterns. {The lagged partial repayment history indicator} has a negative coefficient for staying in good repayment and a positive coefficient for deterioration from good to poor repayment. Larger principal is associated with lower odds of remaining in good repayment, higher odds of deterioration from good to poor repayment, and lower odds of recovery from poor repayment. For monthly repayment, the coefficients indicate higher odds of remaining in good repayment, higher odds of recovery from poor repayment, and lower odds of remaining in poor repayment.

\subsection{Latent Borrower Heterogeneity}
\label{subsec_results_latent_heterogeneity}

Tables \ref{table_full_fe_estimates} to \ref{table_rp_estimates} report the full covariate estimates and frailty standard deviations in order of model complexity. Table \ref{table_full_fe_estimates} gives the fixed-effect estimates, Table \ref{table_ri_estimates} adds the random intercept, and Tables \ref{tab_rl_estimates} and \ref{table_rp_estimates} give the random linear and random piecewise estimates.
\medskip

The random intercept model generally retains the signs and approximate magnitudes of the fixed-effect estimates, with larger differences for some transitions from origin state 2 and a few covariates in models with origin state 3.

The estimated random intercept standard deviations are $1.442$, $1.306$, and $1.127$ for recovery from intermediate repayment $(2,1)$, recovery from poor repayment $(3,1)$, and poor repayment persistence $(3,3)$, respectively. The parametric bootstrap LRTs support nonzero random intercept variance for these three transition-types. They do not support a nonzero random intercept variance for the other transition-types. Each $p$-value is based on $B=1000$ bootstrap replications of the corresponding likelihood ratio test of $H_0:\sigma^2=0$ against $H_1:\sigma^2>0$.
\medskip

The time-dependent variance components in Tables \ref{tab_rl_estimates} and \ref{table_rp_estimates} are not supported by the bootstrap tests. Their covariate estimates also differ more from the fixed-effect estimates than those from the random intercept model, particularly for the age bands and some transitions with origin state 2. Because repayment duration is represented through the frailty structure in these models, the covariate estimates are not directly comparable with the fixed-effect estimates.

\begin{sidewaystable}[!h]
	\centering
	\begin{minipage}{0.98\textheight}
	\centering
	\caption{Fixed-effect LLink estimates by transition-type}
	\label{table_full_fe_estimates}
	\scriptsize
	\renewcommand{\arraystretch}{0.95}
	\setlength{\tabcolsep}{1.3pt}
	\begin{tabular*}{\linewidth}{@{\extracolsep{\fill}}lcccccc@{}}
		\toprule
		Covariate & (1,1) & (1,3) & (2,1) & (2,3) & (3,1) & (3,3) \\
		\midrule
		Main branch & 0.019 (.834) & -0.027 (.794) & 0.428 (.151) & -0.083 (.785) & -0.150 (.434) & 0.194 (.307) \\
		Principal & -2.185 (.003) & 2.604 (.001) & -1.261 (.439) & -0.517 (.747) & -2.730 (.029) & 1.571 (.197) \\
		Age: 26--35 & -0.270 (.524) & 0.201 (.677) & 0.541 (.597) & -0.793 (.446) & 0.931 (.373) & -0.426 (.671) \\
		Age: 36--45 & -0.290 (.493) & 0.145 (.764) & 0.701 (.487) & -0.758 (.459) & 0.846 (.418) & -0.276 (.782) \\
		Age: 46--55 & -0.267 (.528) & 0.144 (.765) & 0.580 (.568) & -0.538 (.602) & 1.081 (.301) & -0.419 (.676) \\
		Age: 56+ & -0.389 (.362) & 0.195 (.688) & 0.908 (.381) & -0.961 (.362) & 0.777 (.460) & -0.375 (.709) \\
		CPI (lag) & 1.789 $(<0.001)$ & -3.160 $(<0.001)$ & -2.028 (.022) & 2.000 (.047) & -4.541 $(<0.001)$ & 4.711 $(<0.001)$ \\
		FX (lag) & 0.344 (.177) & -0.988 (.001) & 0.489 (.405) & 0.026 (.962) & 0.924 (.020) & -1.435 $(<0.001)$ \\
		{Partial repay. (lag)} & -2.524 (.016) & 2.402 $(<0.001)$ & -0.309 (.579) & 1.160 (.035) & -1.167 (.039) & 0.331 (.423) \\
		Long vacation & 1.072 $(<0.001)$ & -1.369 $(<0.001)$ & -0.027 (.922) & 0.590 (.069) & -1.237 $(<0.001)$ & 1.040 $(<0.001)$ \\
		Eid season & 1.068 $(<0.001)$ & -1.378 $(<0.001)$ & -0.442 (.128) & 0.145 (.651) & 0.236 (.326) & -0.249 (.299) \\
		Gender & 0.014 (.890) & 0.055 (.635) & 0.009 (.976) & -0.038 (.902) & 0.288 (.166) & -0.353 (.084) \\
		Group loan & -0.268 (.007) & 0.192 (.095) & 0.060 (.844) & -0.309 (.323) & 0.192 (.365) & -0.228 (.282) \\
		Monthly & 0.544 $(<0.001)$ & 0.150 (.141) & -0.372 (.177) & 0.272 (.328) & 0.925 $(<0.001)$ & -0.765 $(<0.001)$ \\
		Married & 0.107 (.174) & -0.084 (.349) & -0.261 (.286) & -0.083 (.750) & -0.141 (.408) & 0.201 (.231) \\
		Interest rate & 1.388 (.012) & -2.463 $(<0.001)$ & -0.022 (.990) & -1.568 (.375) & 0.941 (.411) & -1.120 (.320) \\
		\bottomrule
	\end{tabular*}
	\par\vspace{2pt}
	{\footnotesize\raggedright Each cell reports the coefficient estimate and its Wald-type $p$-value in parentheses.\par}
	\end{minipage}

	\vspace{4pt}

	\begin{minipage}{0.98\textheight}
	\centering
	\caption{Random intercept LLink estimates by transition-type}
	\label{table_ri_estimates}
	\scriptsize
	\renewcommand{\arraystretch}{0.95}
	\setlength{\tabcolsep}{1.3pt}
	\begin{tabular*}{\linewidth}{@{\extracolsep{\fill}}lcccccc@{}}
		\toprule
		Covariate & (1,1) & (1,3) & (2,1) & (2,3) & (3,1) & (3,3) \\
		\midrule
		Main branch & 0.019 (.83) & -0.027 (.79) & 0.575 (.18) & -0.082 (.79) & -0.099 (.71) & 0.165 (.51) \\
		Principal & -2.189 $(<0.01)$ & 2.594 $(<0.01)$ & -1.502 (.49) & -0.513 (.75) & -3.888 (.02) & 2.221 (.15) \\
		Age: 26--35 & -0.268 (.53) & 0.199 (.68) & 0.834 (.58) & -0.793 (.45) & 1.165 (.36) & -0.613 (.62) \\
		Age: 36--45 & -0.288 (.50) & 0.143 (.77) & 1.019 (.49) & -0.759 (.46) & 1.134 (.38) & -0.480 (.69) \\
		Age: 46--55 & -0.266 (.53) & 0.143 (.77) & 0.883 (.56) & -0.539 (.60) & 1.400 (.28) & -0.636 (.60) \\
		Age: 56+ & -0.387 (.37) & 0.194 (.69) & 1.358 (.38) & -0.962 (.36) & 1.048 (.42) & -0.617 (.62) \\
		CPI (lag) & 1.789 $(<0.01)$ & -3.160 $(<0.01)$ & -2.514 (.04) & 2.001 (.05) & -5.485 $(<0.01)$ & 5.315 $(<0.01)$ \\
		FX (lag) & 0.344 (.18) & -0.988 $(<0.01)$ & 0.410 (.59) & 0.025 (.96) & 1.221 (.01) & -1.710 $(<0.01)$ \\
		{Partial repay. (lag)} & -2.507 (.02) & 2.402 $(<0.01)$ & 0.630 (.43) & 1.160 (.04) & -1.323 (.06) & 0.427 (.41) \\
		Long vacation & 1.072 $(<0.01)$ & -1.369 $(<0.01)$ & -0.041 (.91) & 0.590 (.07) & -1.449 $(<0.01)$ & 1.133 $(<0.01)$ \\
		Eid season & 1.068 $(<0.01)$ & -1.378 $(<0.01)$ & -0.719 (.08) & 0.145 (.65) & 0.436 (.14) & -0.425 (.13) \\
		Gender & 0.014 (.89) & 0.054 (.64) & 0.124 (.77) & -0.038 (.90) & 0.373 (.19) & -0.424 (.11) \\
		Group loan & -0.268 (.01) & 0.192 (.10) & 0.129 (.76) & -0.308 (.32) & 0.137 (.63) & -0.226 (.41) \\
		Monthly & 0.544 $(<0.01)$ & 0.151 (.14) & -0.524 (.18) & 0.272 (.33) & 1.156 $(<0.01)$ & -0.956 $(<0.01)$ \\
		Married & 0.107 (.18) & -0.084 (.35) & -0.403 (.25) & -0.083 (.75) & -0.133 (.57) & 0.220 (.32) \\
		Interest rate & 1.387 (.01) & -2.467 $(<0.01)$ & 1.109 (.65) & -1.562 (.38) & 1.237 (.43) & -1.500 (.31) \\
		$\sigma_u$ & 0.051 (.41) & 0.014 (1.00) & 1.442 $(<0.01)$ & 0.005 (1.00) & 1.306 $(<0.01)$ & 1.127 $(<0.01)$ \\
		\bottomrule
	\end{tabular*}
	\par\vspace{2pt}
	{\footnotesize\raggedright Covariate rows report the coefficient estimate ($p$-value). The $\sigma_u$ row reports the estimated random intercept standard deviation and the bootstrap likelihood ratio $p$-value for the corresponding variance component.\par}
	\end{minipage}
\end{sidewaystable}

\begin{sidewaystable}[!h]
	\centering
	\begin{minipage}{0.98\textheight}
	\centering
	\caption{Random linear frailty LLink estimates by transition-type}
	\label{tab_rl_estimates}
	\scriptsize
	\renewcommand{\arraystretch}{0.95}
	\setlength{\tabcolsep}{1.3pt}
	\begin{tabular*}{\linewidth}{@{\extracolsep{\fill}}lcccccc@{}}
		\toprule
		Covariate & (1,1) & (1,3) & (2,1) & (2,3) & (3,1) & (3,3) \\
		\midrule
		Main branch & 0.021 (.82) & -0.002 (.99) & 0.870 (.11) & -0.063 (.83) & 0.058 (.82) & -0.047 (.84) \\
		Principal & -2.608 $(<0.01)$ & 3.717 $(<0.01)$ & -2.785 (.35) & -0.569 (.74) & -3.514 (.03) & 1.730 (.26) \\
		Age: 26--35 & -0.972 $(<0.01)$ & 0.843 $(<0.01)$ & 2.524 (.01) & -2.687 $(<0.01)$ & 4.106 $(<0.01)$ & -4.126 $(<0.01)$ \\
		Age: 36--45 & -0.916 $(<0.01)$ & 0.632 (.02) & 2.615 (.01) & -2.638 $(<0.01)$ & 4.082 $(<0.01)$ & -4.013 $(<0.01)$ \\
		Age: 46--55 & -0.941 $(<0.01)$ & 0.762 (.01) & 2.397 (.02) & -2.427 $(<0.01)$ & 4.397 $(<0.01)$ & -4.210 $(<0.01)$ \\
		Age: 56+ & -1.112 $(<0.01)$ & 0.875 $(<0.01)$ & 2.988 (.01) & -2.841 $(<0.01)$ & 3.899 $(<0.01)$ & -4.035 $(<0.01)$ \\
		CPI (lag) & 1.270 $(<0.01)$ & -2.468 $(<0.01)$ & -5.221 $(<0.01)$ & 3.804 $(<0.01)$ & -7.527 $(<0.01)$ & 7.113 $(<0.01)$ \\
		FX (lag) & -0.660 (.01) & 0.830 (.01) & 1.082 (.25) & -0.115 (.80) & 1.350 $(<0.01)$ & -1.699 $(<0.01)$ \\
		{Partial repay. (lag)} & -2.561 (.02) & 3.382 $(<0.01)$ & 0.307 (.75) & 1.440 (.01) & -1.296 (.06) & 0.409 (.41) \\
		Long vacation & 0.958 $(<0.01)$ & -0.904 $(<0.01)$ & 0.311 (.38) & 0.660 (.01) & -1.339 $(<0.01)$ & 0.981 $(<0.01)$ \\
		Eid season & 0.900 $(<0.01)$ & -1.026 $(<0.01)$ & 0.094 (.78) & -0.294 (.21) & 0.494 (.03) & -0.524 (.02) \\
		Gender & -0.031 (.77) & 0.083 (.52) & 0.366 (.43) & -0.080 (.79) & 0.469 (.09) & -0.531 (.04) \\
		Group loan & -0.262 (.01) & 0.062 (.62) & 0.254 (.59) & -0.542 (.06) & 0.316 (.25) & -0.398 (.12) \\
		Monthly & 0.562 $(<0.01)$ & -0.005 (.96) & -0.815 (.09) & 0.247 (.35) & 1.378 $(<0.01)$ & -1.125 $(<0.01)$ \\
		Married & 0.230 (.01) & -0.344 $(<0.01)$ & -0.081 (.83) & -0.339 (.17) & 0.055 (.81) & -0.032 (.88) \\
		Interest rate & -0.447 (.40) & -0.528 (.43) & 2.163 (.40) & -1.265 (.42) & 0.673 (.62) & -0.547 (.66) \\
		$\sigma_a$ & 0.108 (1.00) & 0.251 (1.00) & 0.508 (1.00) & 0.002 (1.00) & 0.001 (1.00) & 0.001 (1.00) \\
		$\sigma_b$ & 0.006 (1.00) & 0.002 (1.00) & 0.001 (1.00) & 0.001 (1.00) & 1.320 (1.00) & 1.116 (1.00) \\
		\bottomrule
	\end{tabular*}
	\par\vspace{2pt}
	{\footnotesize\raggedright Covariate cells report the coefficient estimate ($p$-value). The $\sigma_a$ and $\sigma_b$ rows report estimated standard deviations and bootstrap likelihood ratio $p$-values for the corresponding variance components.\par}
	\end{minipage}

	\vspace{4pt}

	\begin{minipage}{0.98\textheight}
	\centering
	\caption{Random piecewise frailty LLink estimates by transition-type}
	\label{table_rp_estimates}
	\scriptsize
	\renewcommand{\arraystretch}{0.95}
	\setlength{\tabcolsep}{1.3pt}
	\begin{tabular*}{\linewidth}{@{\extracolsep{\fill}}lcccccc@{}}
		\toprule
		Covariate & (1,1) & (1,3) & (2,1) & (2,3) & (3,1) & (3,3) \\
		\midrule
		Main branch & -0.014 (.88) & -0.015 (.88) & 0.414 (.15) & -0.063 (.83) & -0.031 (.87) & 0.009 (.96) \\
		Principal & -2.557 $(<0.01)$ & 3.031 $(<0.01)$ & -1.418 (.35) & -0.687 (.69) & -2.481 (.05) & 1.178 (.32) \\
		Age: 26--35 & -1.291 $(<0.01)$ & 0.848 $(<0.01)$ & 2.147 $(<0.01)$ & -2.401 $(<0.01)$ & 3.117 $(<0.01)$ & -3.380 $(<0.01)$ \\
		Age: 36--45 & -1.240 $(<0.01)$ & 0.696 $(<0.01)$ & 2.242 $(<0.01)$ & -2.333 $(<0.01)$ & 3.046 $(<0.01)$ & -3.235 $(<0.01)$ \\
		Age: 46--55 & -1.237 $(<0.01)$ & 0.739 $(<0.01)$ & 2.142 $(<0.01)$ & -2.133 $(<0.01)$ & 3.303 $(<0.01)$ & -3.398 $(<0.01)$ \\
		Age: 56+ & -1.389 $(<0.01)$ & 0.847 $(<0.01)$ & 2.500 $(<0.01)$ & -2.545 $(<0.01)$ & 2.906 $(<0.01)$ & -3.243 $(<0.01)$ \\
		CPI (lag) & 1.568 $(<0.01)$ & -2.800 $(<0.01)$ & -3.557 $(<0.01)$ & 3.378 $(<0.01)$ & -5.869 $(<0.01)$ & 5.886 $(<0.01)$ \\
		FX (lag) & -0.038 (.88) & 0.083 (.78) & 0.914 (.07) & -0.073 (.87) & 1.187 $(<0.01)$ & -1.581 $(<0.01)$ \\
		{Partial repay. (lag)} & -2.642 (.01) & 2.629 $(<0.01)$ & -0.737 (.18) & 1.421 (.01) & -1.138 (.05) & 0.288 (.48) \\
		Long vacation & 0.986 $(<0.01)$ & -0.846 $(<0.01)$ & -0.137 (.52) & 0.682 (.01) & -1.153 $(<0.01)$ & 0.933 $(<0.01)$ \\
		Eid season & 0.964 $(<0.01)$ & -1.026 $(<0.01)$ & 0.195 (.37) & -0.337 (.16) & 0.380 (.05) & -0.449 (.02) \\
		Gender & -0.048 (.62) & 0.111 (.31) & 0.037 (.89) & -0.077 (.79) & 0.344 (.09) & -0.428 (.03) \\
		Group loan & -0.286 $(<0.01)$ & 0.137 (.20) & 0.231 (.42) & -0.496 (.09) & 0.298 (.14) & -0.348 (.08) \\
		Monthly & 0.516 $(<0.01)$ & 0.104 (.28) & -0.394 (.14) & 0.253 (.34) & 1.048 $(<0.01)$ & -0.868 $(<0.01)$ \\
		Married & 0.166 (.03) & -0.220 (.01) & -0.043 (.85) & -0.313 (.21) & -0.027 (.87) & 0.045 (.78) \\
		Interest rate & 0.273 (.60) & -0.883 (.13) & 0.405 (.79) & -1.073 (.49) & 0.855 (.40) & -0.837 (.40) \\
		$\sigma_1$ & 0.248 (1.00) & 0.054 (1.00) & 0.000 (1.00) & 0.511 (1.00) & 0.227 (1.00) & 0.171 (1.00) \\
		$\sigma_2$ & 0.003 (1.00) & 1.132 (1.00) & 0.261 (1.00) & 0.001 (1.00) & 0.001 (1.00) & 0.004 (1.00) \\
		$\sigma_3$ & 0.840 (1.00) & 0.895 (1.00) & 0.001 (1.00) & 0.001 (1.00) & 0.342 (1.00) & 0.367 (1.00) \\
		\bottomrule
	\end{tabular*}
	\par\vspace{2pt}
	{\footnotesize\raggedright Covariate rows report the coefficient estimate ($p$-value). The $\sigma_1$, $\sigma_2$, and $\sigma_3$ rows report estimated standard deviations and bootstrap likelihood-ratio $p$-values for the corresponding repayment period-specific variance components.\par}
	\end{minipage}
\end{sidewaystable}

\clearpage

\subsection{Prediction performance}
\label{subsec_results_prediction_checks}

Table \ref{tab:full-prediction-performance} compares one-step prediction performance from the fixed-effect, random intercept, random linear, and random piecewise LLink models with Random Forest (RF) and KTBoost for each transition-type $(h,j)$ \citep{breiman2001random,sigrist2021ktboost}. We also report the area under the receiver operating characteristic curve (ROC-AUC) \citep{hanley1982meaning} and average precision (AP) \citep{saito2015precision}.

All models use the same borrower-level 80/20 training and test split, with all observations for a borrower assigned to either the training set or the test set. Predictions from the frailty models are marginal probabilities obtained using GHQ.
\medskip

RF and KTBoost have higher values for the four transition-types with origin state 1 or origin state 3. The comparisons for the two transition-types with origin state 2 are less consistent. For $(2,1)$, the fixed-effect LLink model has the highest AP, whereas RF has the highest ROC-AUC. For $(2,3)$, RF and the LLink models have almost the same ROC-AUC and AP. With only 132 test rows for each comparison with origin state 2, we cannot draw firm conclusions about which model performs better for these transition-types.
	\medskip
	
Within the LLink models, the fixed-effect and random intercept models perform similarly across transition-types. The random intercept provides evidence of remaining borrower heterogeneity but does not materially improve one-step prediction. The random linear and random piecewise models also do not consistently improve test-set ROC-AUC or AP. In this paper, these models are used mainly to examine whether unobserved borrower differences change over repayment time.

\begin{table}[!h]
	\centering
	\caption{{One-step binary prediction performance across models and transition-types}}
	\label{tab:full-prediction-performance}
	\scriptsize
	\setlength{\tabcolsep}{5pt}
	\renewcommand{\arraystretch}{1.05}
	\begin{tabular}{@{}lccccccc@{}}
		\toprule
		Transition &
		\shortstack{Event rate\\(\%)} &
		RF &
		KTBoost &
		FE &
		RI &
		Linear &
		Piecewise \\
		\midrule
		\multicolumn{8}{@{}l}{\textit{Metric: ROC-AUC}} \\
		$(1,1)$ & 58.3 & 0.838 & 0.836 & 0.736 & 0.736 & 0.714 & 0.728 \\
		$(1,3)$ & 30.5 & 0.859 & 0.855 & 0.733 & 0.733 & 0.729 & 0.746 \\
		$(2,1)$ & 46.2 & 0.825 & 0.790 & 0.801 & 0.794 & 0.742 & 0.743 \\
		$(2,3)$ & 40.2 & 0.815 & 0.771 & 0.814 & 0.815 & 0.753 & 0.758 \\
		$(3,1)$ & 62.5 & 0.854 & 0.859 & 0.829 & 0.829 & 0.818 & 0.821 \\
		$(3,3)$ & 32.2 & 0.855 & 0.867 & 0.821 & 0.822 & 0.811 & 0.800 \\
		\midrule
		\multicolumn{8}{@{}l}{\textit{Metric: Average precision}} \\
		$(1,1)$ & 58.3 & 0.854 & 0.852 & 0.739 & 0.739 & 0.714 & 0.733 \\
		$(1,3)$ & 30.5 & 0.802 & 0.806 & 0.682 & 0.682 & 0.652 & 0.677 \\
		$(2,1)$ & 46.2 & 0.813 & 0.804 & 0.828 & 0.819 & 0.789 & 0.772 \\
		$(2,3)$ & 40.2 & 0.739 & 0.631 & 0.734 & 0.734 & 0.615 & 0.626 \\
		$(3,1)$ & 62.5 & 0.918 & 0.926 & 0.903 & 0.903 & 0.897 & 0.901 \\
		$(3,3)$ & 32.2 & 0.692 & 0.716 & 0.588 & 0.587 & 0.615 & 0.593 \\
		\bottomrule
	\end{tabular}
	
	\par\vspace{2pt}
	\footnotesize\textit{Note:} {FE and RI represent the fixed-effect and random intercept LLink models; Linear and Piecewise represent the random linear and random piecewise LLink models.}\par
\end{table}
\medskip

We also compare the Djeundje and Crook (D\&C) multistate classification rule \citep{djeundje2018incorporating} with the Optimised Matthews Correlation Coefficient (OMCC) rule proposed in this paper. Further details of the two rules and prediction measures are given in Appendix~\ref{app_prediction_measures}. 
		
{The comparison uses transition probabilities estimated by RF. For each borrower and repayment period, the one-step probabilities form a $3\times3$ transition matrix, whose $h$th row contains the probabilities of moving from origin state $h$ to each destination state. For $r<s$, the probabilities are obtained by multiplying these matrices in chronological order from period $r+1$ to period $s$. Let $\tilde q_{i,hj}(r,s)$ denote the predicted probability that borrower $i$ is in state $j$ at period $s$, given that the borrower was in state $h$ at period $r$. Tables~\ref{table_threshold_accuracy} and \ref{table_threshold_classification} report results for horizons $(r,s)=(1,2)$ and $(r,s)=(2,4)$.} 
\medskip 

Given the origin-specific cut-offs $\boldsymbol a_h=(a_{h1},a_{h2},a_{h3})^{\top}$, both rules classify the state at period $s$ as
\begin{align}\label{eq:main-threshold-rule}
\widehat S_i(s;\boldsymbol a_h)= \arg\max_{j\in\{1,2,3\}} \{\tilde q_{i,hj}(r,s)-a_{hj}\}.
\end{align}
D\&C selects the cut-offs that maximise classification accuracy, whereas OMCC selects the cut-offs that maximise the multiclass Matthews correlation coefficient (MCC) \citep{matthews1975comparison,gorodkin2004comparing}. The borrower-level training and test split is kept fixed. The RF models are fitted once on the training set, and their estimated probabilities remain fixed across replications. In each of 1000 bootstrap replications, borrowers are sampled with replacement from the training set, the cut-offs are selected using the bootstrap sample, and the resulting cut-offs are evaluated on the fixed test set. Tables \ref{table_threshold_accuracy} and \ref{table_threshold_classification} report the mean of each performance measure across the 1000 replications, together with the standard deviation of accuracy.

\begin{table}[!htbp]
	\centering
	\caption{Bootstrap accuracy and recall for delinquency and state 1 under the D\&C and OMCC threshold rules}
	\label{table_threshold_accuracy}
	\scriptsize
	\setlength{\tabcolsep}{3pt}
	\begin{tabular}{lrlrrrr}
		\toprule
			\makecell{Prediction\\horizon} & Origin & Rule &
			\makecell{Mean\\accuracy} & \makecell{Accuracy\\SD} &
			\makecell{Recall\\(delinquency)} & \makecell{Recall\\(state 1)}\\
		\midrule
		$1\to 2$ & 1 & D\&C  & 90.0 & 0.8 & 89.8 & 94.4 \\
		$1\to 2$ & 1 & OMCC  & 89.9 & 0.8 & 89.8 & 94.4 \\
		$1\to 2$ & 2 & D\&C  & 82.0 & 4.8 & 34.5 & 93.9 \\
		$1\to 2$ & 2 & OMCC  & 79.3 & 5.6 & 38.7 & 90.6 \\
		$1\to 2$ & 3 & D\&C  & 80.2 & 1.2 & 16.7 & 93.6 \\
		$1\to 2$ & 3 & OMCC  & 79.8 & 1.3 & 16.7 & 93.1 \\
		$2\to 4$ & 1 & D\&C  & 56.3 & 1.6 & 49.4 & 71.3 \\
		$2\to 4$ & 1 & OMCC  & 55.9 & 2.2 & 51.7 & 69.8 \\
		$2\to 4$ & 2 & D\&C  & 41.3 & 10.6 & 89.3 & 46.2 \\
		$2\to 4$ & 2 & OMCC  & 40.4 & 10.9 & 90.0 & 45.2 \\
		$2\to 4$ & 3 & D\&C  & 65.0 & 1.4 & 44.4 & 80.8 \\
		$2\to 4$ & 3 & OMCC  & 64.3 & 1.6 & 47.6 & 78.7 \\
		\bottomrule
	\end{tabular}
	\par\vspace{2pt}
	{\footnotesize\raggedright{Mean accuracy, the standard deviation of accuracy, and recall are reported as percentages.}\par}
\end{table}

\begin{table}[!htbp]
	\centering
	\caption{{Bootstrap balanced accuracy, MCC, and precision under the D\&C and OMCC threshold rules}}
	\label{table_threshold_classification}
	\scriptsize
	\setlength{\tabcolsep}{3pt}
	\begin{tabular}{@{}lcccccc@{}}
		\toprule
		\makecell{{Prediction}\\{horizon}} & Origin & Rule & \makecell{Balanced\\accuracy} & MCC & \makecell{Precision\\(delinquency)} & \makecell{Precision\\(state 1)} \\
		\midrule
		$1\to 2$ & 1 & D\&C & 83.0 & 0.805 & 96.5 & 84.3 \\
		$1\to 2$ & 1 & OMCC & 83.1 & 0.804 & 96.5 & 84.3 \\
		$1\to 2$ & 2 & D\&C & 55.3 & 0.273 & 62.9 & 88.8 \\
		$1\to 2$ & 2 & OMCC & 53.8 & 0.235 & 52.3 & 89.1 \\
		$1\to 2$ & 3 & D\&C & 46.8 & 0.044 & 31.1 & 87.1 \\
		$1\to 2$ & 3 & OMCC & 46.6 & 0.037 & 29.6 & 87.0 \\
		$2\to 4$ & 1 & D\&C & 44.6 & 0.191 & 52.4 & 69.1 \\
		$2\to 4$ & 1 & OMCC & 45.3 & 0.198 & 52.4 & 70.0 \\
		$2\to 4$ & 2 & D\&C & 47.1 & 0.208 & 47.9 & 94.4 \\
		$2\to 4$ & 2 & OMCC & 45.8 & 0.199 & 48.8 & 95.1 \\
		$2\to 4$ & 3 & D\&C & 44.2 & 0.237 & 54.4 & 73.9 \\
		$2\to 4$ & 3 & OMCC & 44.4 & 0.239 & 53.7 & 74.6 \\
		\bottomrule
	\end{tabular}
	\par\vspace{2pt}
	{\footnotesize\raggedright Balanced accuracy and precision are reported as percentages; MCC ranges from $-1$ to $1$.\par}
\end{table}
\medskip

Over the $1\to2$ horizon, the two rules perform almost identically for borrowers starting in state 1. For those starting in state 2, OMCC has higher delinquency recall ($38.7\%$ versus $34.5\%$), whereas D\&C has higher overall accuracy ($82.0\%$ versus $79.3\%$), balanced accuracy ($55.3\%$ versus $53.8\%$), MCC ($0.273$ versus $0.235$), and delinquency precision ($62.9\%$ versus $52.3\%$). Similar trade-offs appear over the longer horizon, and neither rule performs best across all measures, origin states, and horizons.

\section{Discussion and Conclusion}
\label{section_discussion}

During the short life of a microfinance loan, a borrower can repay close to the scheduled amount, deteriorate into poor repayment, stay in arrears, or recover. Under a binary classification, where both states 2 and 3 are classified as delinquent, a transition from good to intermediate repayment and a direct move from good to poor repayment are treated as the same movement. The same issue occurs with recovery from intermediate or poor repayment to good repayment. The three-state formulation keeps these movements separate and estimates covariate associations for each transition-type, while allowing the baseline transition probabilities to vary over repayment duration.
	\medskip
	
The estimates show that the associations of loan terms, repayment history, calendar indicators, and macroeconomic variables differ across the modelled transition-types, including between persistence, deterioration, and recovery. The socio-temporal indicators mark the timing of calendar periods, while the macroeconomic variables represent the macroeconomic context observed before the transition. {These variables do not measure the household or business conditions through which the associations may occur.}
	\medskip

The bootstrap likelihood ratio tests provide evidence of remaining borrower-level variation in recovery from intermediate repayment, recovery from poor repayment, and poor-repayment persistence. For these transition-types, the random intercept captures differences between borrowers after accounting for the fixed-effect covariates and repayment duration.
	
However, accounting for this variation does not improve one-step prediction. The fixed-effect and random intercept LLink models have very similar ROC-AUC and average precision values, while RF and KTBoost perform better for several transition-types. The D\&C and OMCC comparison shows that neither threshold rule performs best across all reported measures, origin states, and prediction horizons.
	\medskip
	
Our study uses the loanbook records from one microfinance institution over a short observation window, covering products with short repayment histories. These short repayment histories are a feature of the loan products observed here, but they limit the within-borrower information available for estimating transition patterns. {The calendar and macroeconomic covariates may also capture other unrecorded time-related conditions, so their estimated effects should not be interpreted as causal effects.} The intermediate state is less common in the data, so the recovery and deterioration estimates from that state, together with the related prediction comparisons, {rely} on fewer observations. The prediction and threshold checks were conducted on a test sample from the same loanbook, but their performance with records from other microfinance institutions, and their effects when used to guide staff intervention, require further study.
	\medskip
	
The next step is to assess the transition patterns using longer repayment histories from other microfinance institutions, with more observations from the intermediate state. They could also provide more transitions from the intermediate state, which would allow the three destination probabilities to be estimated jointly. Richer household and business data would help explain the calendar and macroeconomic associations. The practical value of D\&C and OMCC also depends on prospective evidence about the costs and outcomes of using the rules to guide staff intervention.

\clearpage
\appendix 

\section{Variance component tests}
\label{app_variance_tests}

For each {transition-type} $(h,j)$, the parametric bootstrap likelihood-ratio test compares a full model with the corresponding reduced model \citep{davison1997bootstrap}. The observed statistic is
$$
\Lambda_{hj}=2\big(\ell_{hj}^{full}-\ell_{hj}^{reduced}\big).
$$
For each bootstrap replication, data are generated under the reduced model, the reduced and full models are refitted, and the bootstrap statistic $\Lambda_{hj}^{(b)}$ is computed. The bootstrap $p$-value is
$$
p_{hj}^{boot}=\frac{1}{B}\sum_{b=1}^{B} {\mathbbm{1}_{\{\Lambda_{hj}^{(b)}\geq \Lambda_{hj}\}}},
\qquad B=1000.
$$
{For the random intercept test, the reduced model is the fixed-effect LLink model and the full model adds $u_{i,hj}\sim N(0,\sigma^2_{u,hj})$. For the random linear model, $\sigma^2_{a,hj}=0$ is tested while keeping $\sigma^2_{b,hj}$ in the reduced model, and $\sigma^2_{b,hj}=0$ is tested while retaining $\sigma^2_{a,hj}$. For the random piecewise model, each null hypothesis $\sigma^2_{hj,k}=0$ is tested while keeping the other two repayment period-specific variance components. The resulting $p$-values are reported with the variance rows in Tables \ref{table_ri_estimates}-\ref{table_rp_estimates}. The tables report standard deviations, and the tests examine whether the corresponding variances are zero.}

\section{Diagnostics and sensitivity of the state definition}
\label{section_diagnostics}

\begin{figure}[!htbp]
	\centering
	\includegraphics[width=0.88\textwidth]{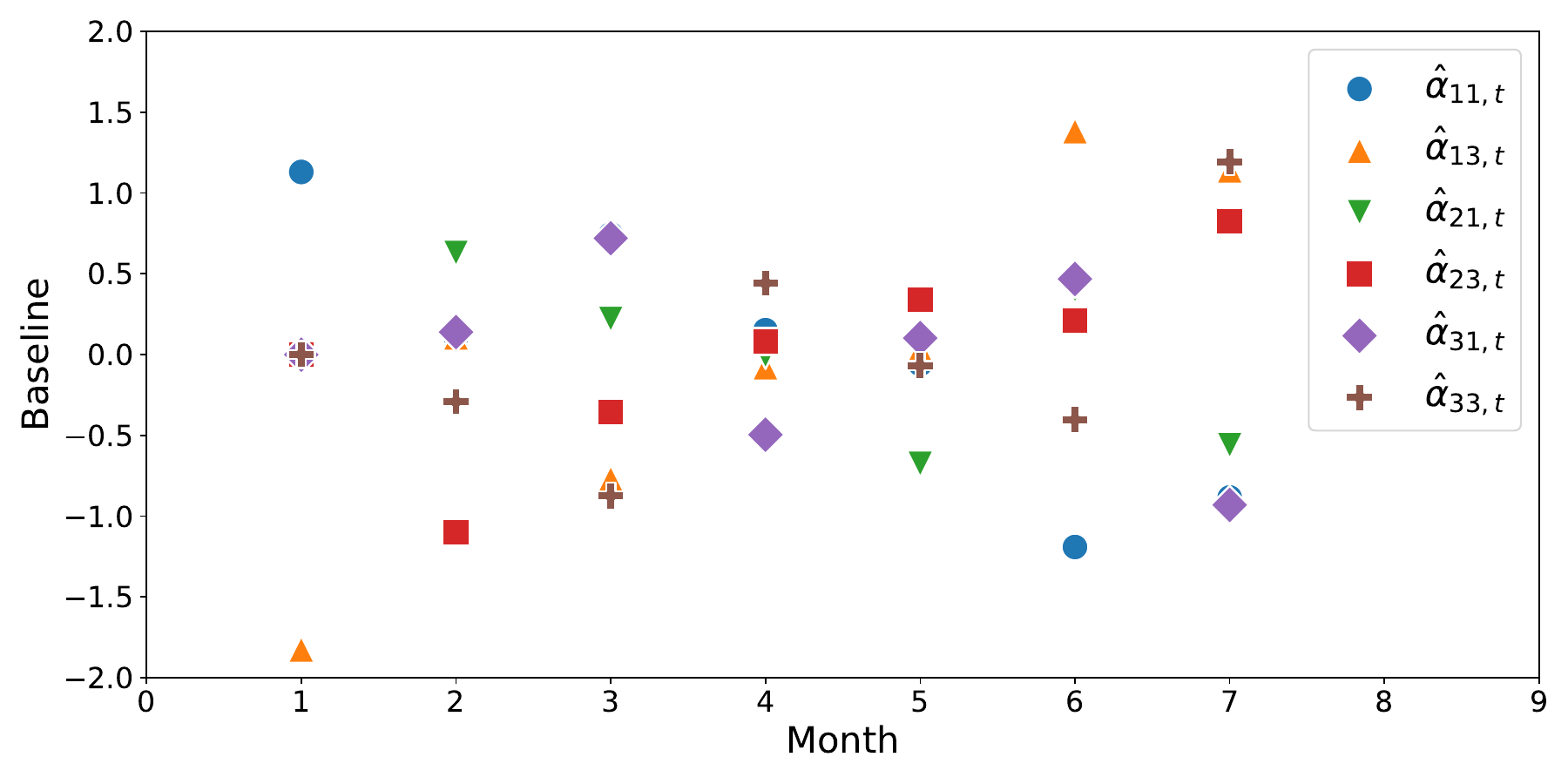}
	\caption{Estimated duration baseline functions for the three-state transition model}
	\label{app_baseline}
\end{figure}

Figure \ref{app_baseline} shows the fitted duration baselines $\hat\alpha_{hj,t}$ from the fixed-effect LLink model. These are transition-specific intercepts on the repayment duration grid. They control for duration effects in $q_{i,hj}(t)$ and do not represent borrower-level effects.

\medskip

\begin{figure}[!htbp]
	\centering
	\includegraphics[width=\textwidth]{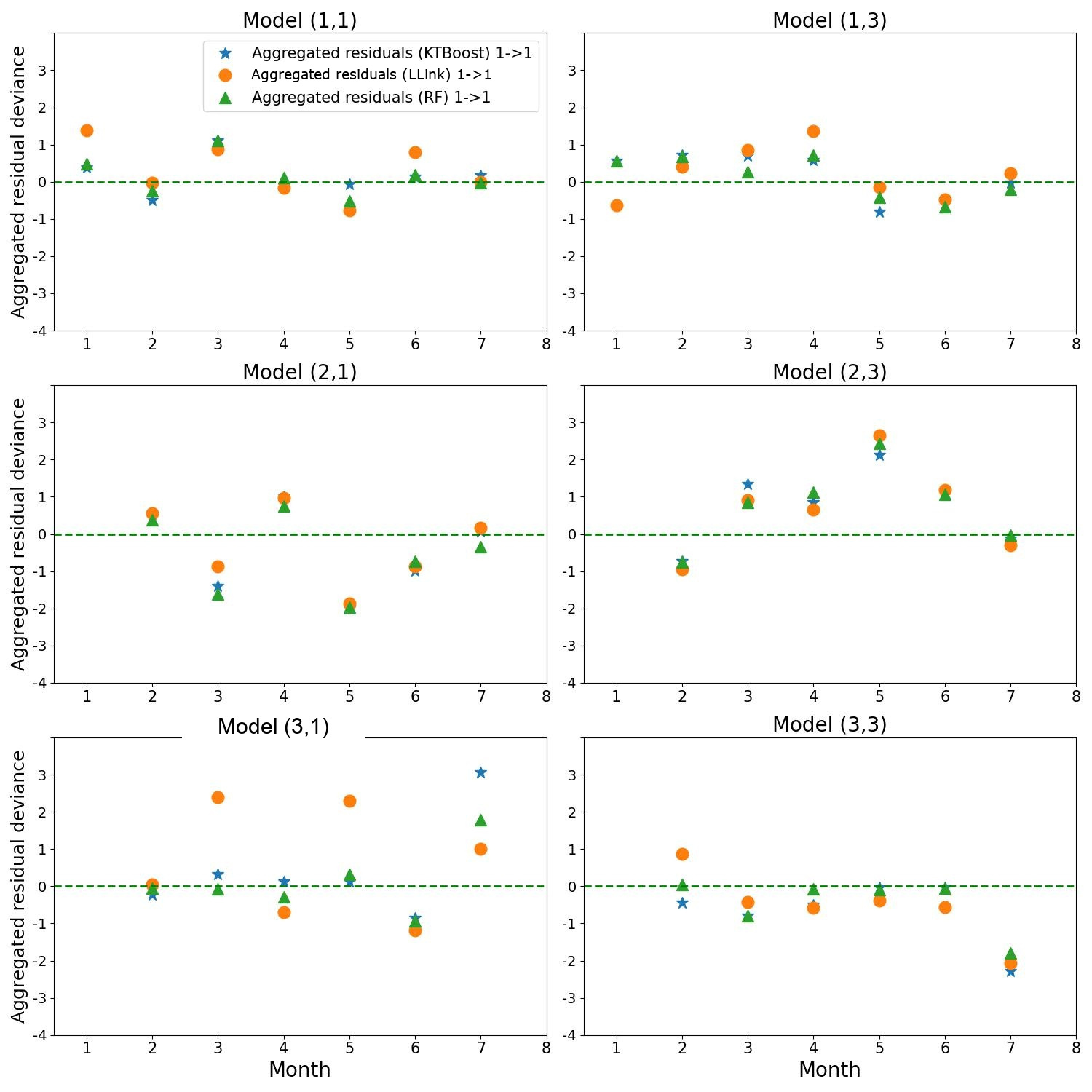}
\caption{Calendar-month aggregated deviance residuals by transition-type and fitted model}
	\label{fig_app_residuals}
\end{figure}

For calendar month $t$, let
$$
\begin{aligned}
N_h(t)=| \mathcal{R}_h(t-1)|, \quad O_{hj}(t) =\sum_{i \in \mathcal{R}_h(t-1)}Y_{i,hj}(t),\quad
E_{hj}(t) =\sum_{i \in \mathcal{R}_h(t-1)}\hat q_{i,hj}(t),
\end{aligned}
$$
where $N_h(t)$ is the number of borrowers at risk of transition before month $t$, $O_{hj}(t)$ and $E_{hj}(t)$ are the number of observed transitions and aggregated estimated probabilities from state $h$ at month $t-1$ to state $j$ at month $t$. The monthly deviance residual is then given as

$$
\begin{aligned}
D_{hj}(t) &=\operatorname{sign}\{O_{hj}(t)-E_{hj}(t)\}
\Bigl[2\Bigl\{ O_{hj}(t)\log\frac{O_{hj}(t)}{E_{hj}(t)}
\\[-1mm]
&\qquad +\big\{N_h(t)-O_{hj}(t)\big\} \log\frac{N_h(t)-O_{hj}(t)}{N_h(t)-E_{hj}(t)} \Bigr\}\Bigr]^{1/2},
\end{aligned}
$$
where $0\log(0/a)=0$. Figure \ref{fig_app_residuals} provides the calendar-month residuals for the LLink, RF, and KTBoost fits. The residuals are calculated from all observed borrower-period pairs in the modelling sample. For most transition-types, the residuals remain close to zero and mostly within $[-2,2]$. The largest departures occur in the origin state 3 panels, including recovery from poor repayment $(3,1)$.

\medskip

\begin{figure}[!h]
	\centering
	\includegraphics[width=15cm,height=8cm]{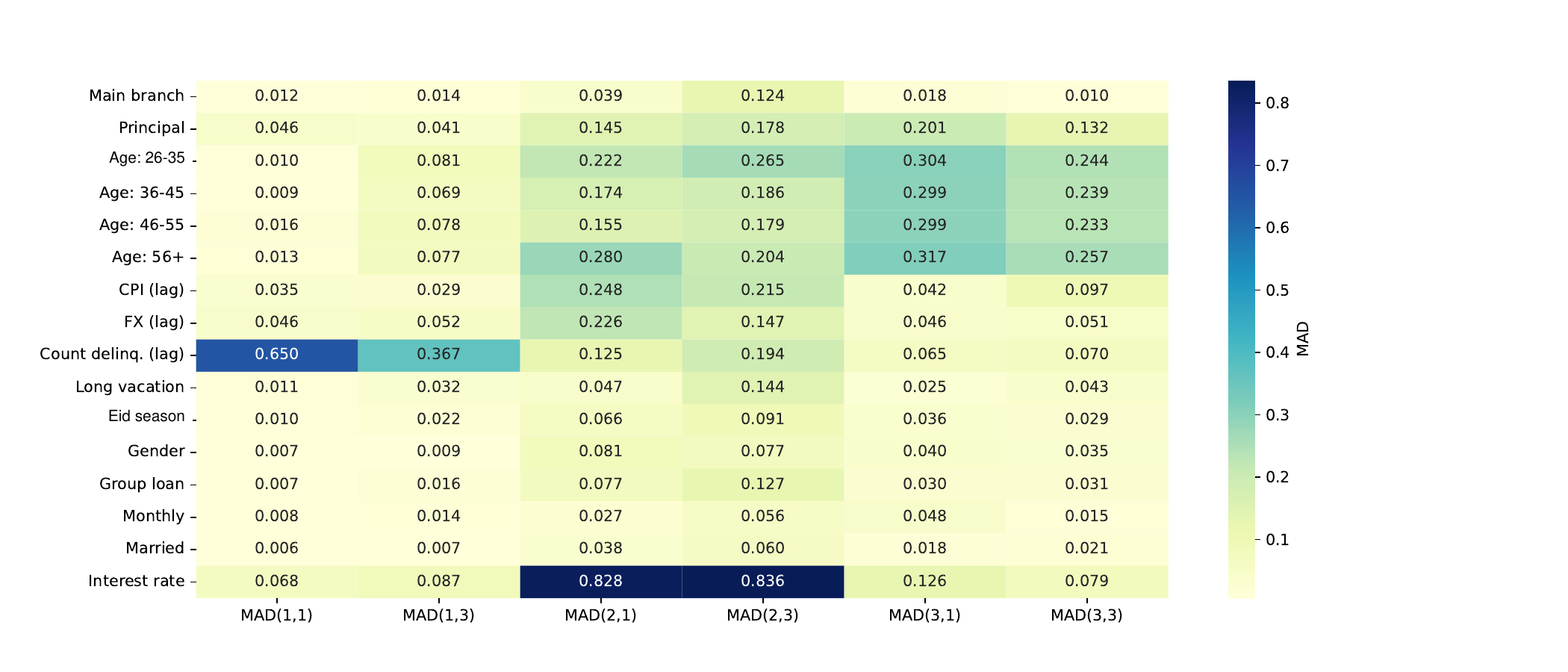}
	\caption{Mean absolute difference of coefficient estimates under alternative state thresholds}
	\label{figure_app_mad}
\end{figure}

To assess the sensitivity of the coefficient estimates to the state definitions, we refit the fixed-effect LLink model for each of the six transition-types using ten alternative threshold pairs. The lower cut-offs lie between $0.575$ and $0.625$, and the upper cut-offs lie between $0.825$ and $0.875$. Let $\mathcal C$ denote the set of ten alternative threshold pairs; the baseline pair is $(0.60,0.82)$. For covariate $k$ and transition-type $(h,j)$, let $\hat\beta^{(c)}_{k,hj}$ denote the coefficient estimate under threshold pair $c\in\mathcal C$, and let $\hat\beta_{k,hj}$ denote the estimate under the baseline thresholds. Figure~\ref{figure_app_mad} reports the mean absolute difference:
$$
\mathrm{MAD}_{k,hj} = 
\frac{1}{|\mathcal C|}
\sum_{c\in\mathcal C}
|\hat\beta^{(c)}_{k,hj}-\hat\beta_{k,hj}|.
$$
The coefficient estimates for transitions from origin states 1 and 3 are more stable than those for transitions from origin state 2. Table~\ref{tab_app_sign_counts} reports sign agreement for the selected covariates shown in the table.

\clearpage

\begin{table}[!h]
	\centering
	\caption{{Sign agreement for selected covariates across alternative threshold definitions}}
	\label{tab_app_sign_counts}
	\scriptsize
	\setlength{\tabcolsep}{3pt}
	\begin{tabular}{lcccccccc}
		\toprule
		Transition & Principal & CPI & FX & {Partial repay.} & Long vac. & Eid & Group loan & Monthly \\
		\midrule
		(1,1) same & 10 & 10 & 10 & 10 & 10 & 10 & 10 & 10 \\
		(1,1) changed & 0 & 0 & 0 & 0 & 0 & 0 & 0 & 0 \\
		(1,3) same & 10 & 10 & 10 & 10 & 10 & 10 & 10 & 10 \\
		(1,3) changed & 0 & 0 & 0 & 0 & 0 & 0 & 0 & 0 \\
		(2,1) same & 10 & 10 & 10 & 10 & 5 & 10 & 3 & 10 \\
		(2,1) changed & 0 & 0 & 0 & 0 & 5 & 0 & 7 & 0 \\
		(2,3) same & 10 & 10 & 4 & 10 & 10 & 9 & 10 & 10 \\
		(2,3) changed & 0 & 0 & 6 & 0 & 0 & 1 & 0 & 0 \\
		(3,1) same & 10 & 10 & 10 & 10 & 10 & 10 & 10 & 10 \\
		(3,1) changed & 0 & 0 & 0 & 0 & 0 & 0 & 0 & 0 \\
		(3,3) same & 10 & 10 & 10 & 10 & 10 & 10 & 10 & 10 \\
		(3,3) changed & 0 & 0 & 0 & 0 & 0 & 0 & 0 & 0 \\
		\bottomrule
	\end{tabular}
	\par\vspace{2pt}
	{\footnotesize\raggedright Counts are out of ten alternative threshold refits. ``Same'' means that the estimate has the same sign as the baseline fixed-effect estimate. ``Changed'' means that its sign is reversed.\par}
\end{table}

\clearpage
\section{Multistate classification rules}
\label{app_prediction_measures}

Both D\&C and OMCC use the classification rule in Equation \eqref{eq:main-threshold-rule} and the same candidate cut-offs. Fix an origin state $h$ at time $r$ and a prediction time $s>r$. Let $N_h(r,s)$ be the number of borrowers with $S_i(r)=h$ for whom $S_i(s)$ and $\big(\tilde q_{i,hj}(r,s)\big)_{j\in\{1,2,3\}}$ are available; all sums below use these borrowers.

Consider $\widehat S_i(s;\boldsymbol a_h)$ from Equation~\eqref{eq:main-threshold-rule}. Let $n_{h,k\ell}(\boldsymbol a_h)$ be the number of borrowers observed in destination state $k$ and classified in destination state $\ell$, where $k,\ell\in\{1,2,3\}$ and
$$
n_{h,k\ell}(\boldsymbol a_h) := \sum_{i:S_i(r)=h} \mathbbm{1}_{\{S_i(s)=k,\;\widehat S_i(s;\boldsymbol a_h)=\ell\}}.
$$
In the expressions below, let $n_{h,k\ell}=n_{h,k\ell}(\boldsymbol a_h)$, with the dependence on $\boldsymbol a_h$ understood, and let $n_{h,k\cdot}=\sum_{\ell=1}^3n_{h,k\ell}$ and $n_{h,\cdot\ell}=\sum_{k=1}^3n_{h,k\ell}$. D\&C selects the threshold vector that maximises
$$
\operatorname{Accuracy}_h(\boldsymbol a_h)
= \frac{1}{N_h(r,s)}\sum_{k=1}^3n_{h,kk},
$$
whereas OMCC selects the threshold vector that maximises
$$
\operatorname{MCC}_h(\boldsymbol a_h) = \frac{N_h(r,s)\sum_{k=1}^3n_{h,kk}-\sum_{k=1}^3n_{h,k\cdot}n_{h,\cdot k}} {\big[\big(N_h(r,s)^2-\sum_{k=1}^3n_{h,k\cdot}^2\big) \big(N_h(r,s)^2-\sum_{k=1}^3n_{h,\cdot k}^2\big)\big]^{1/2}}.
$$

Accuracy is the proportion classified correctly, and balanced accuracy is the average recall across the destination states represented in the test set. 

For recall and precision, define the observed and predicted indicators for delinquency and state 1 by

\begin{align*}
D_i(s)&:=\mathbbm{1}_{\{S_i(s)\in\{2,3\}\}}, & \widehat D_i(s;\boldsymbol a_h)&:=\mathbbm{1}_{\{\widehat S_i(s;\boldsymbol a_h)\in\{2,3\}\}},\\
G_i(s)&:=\mathbbm{1}_{\{S_i(s)=1\}}, & \widehat G_i(s;\boldsymbol a_h)&:=\mathbbm{1}_{\{\widehat S_i(s;\boldsymbol a_h)=1\}}.
\end{align*}

Let $(C_i(s),\widehat C_i(s;\boldsymbol a_h))$ denote either $(D_i(s),\widehat D_i(s;\boldsymbol a_h))$ for delinquency or $(G_i(s),\widehat G_i(s;\boldsymbol a_h))$ for state 1. Then
\begin{align*}
	\operatorname{Recall}_h(C;\boldsymbol a_h)
	&=	\frac{\sum_{i:S_i(r)=h} \mathbbm{1}_{\{C_i(s)=1,\widehat C_i(s;\boldsymbol a_h)=1\}}} {\sum_{i:S_i(r)=h} \mathbbm{1}_{\{C_i(s)=1\}}},\\	\operatorname{Precision}_h(C;\boldsymbol a_h)
		&= \frac{\sum_{i:S_i(r)=h} \mathbbm{1}_{\{C_i(s)=1,\widehat C_i(s;\boldsymbol a_h)=1\}}}	{\sum_{i:S_i(r)=h} \mathbbm{1}_{\{\widehat C_i(s;\boldsymbol a_h)=1\}}}.
\end{align*}
When $C=D$, recall and precision combine destination states 2 and 3; when $C=G$, they refer to destination state 1. Accuracy, balanced accuracy, and MCC use all three destination states. Tables~\ref{table_threshold_accuracy} and \ref{table_threshold_classification} report means across the 1000 bootstrap replications.

\section*{Acknowledgements}
All authors thank Shiela Azuntaba for providing the resources and real-world insights that were essential to this study.

\section*{Funding}
Olivier Menoukeu Pamen acknowledges the funding provided by the Alexander von Humboldt Foundation, under the programme financed by the German Federal Ministry of Education and Research entitled German Research Chair No 01DG15010.

\section*{Data availability}

The data obtained from the microfinance institution are confidential and cannot be shared under the data-sharing agreement. Related code using synthetic data is available in the \href{https://github.com/cedricak/Modeling-credit-risk-delinquency-in-MFI/blob/main/Chapter_2_codes_final.ipynb}{GitHub repository}. The notebook reproduces the separate EM-GHQ simulation study reported in \citet{koffi2026dependence} and does not use the confidential data analysed in this paper.

\section*{Conflict of interest} The authors declared no conflict of interest.

\bibliography{bibliography}

\end{document}